*Article*

# Venus Life Finder Habitability Mission: Motivation, Science Objectives, and Instrumentation

**Sara Seager** [1,2,3,*], **Janusz J. Petkowski** [1], **Christopher E. Carr** [4], **Sarag J. Saikia** [5], **Rachana Agrawal** [1,6], **Weston P. Buchanan** [6], **David H. Grinspoon** [7], **Monika U. Weber** [8], **Pete Klupar** [9], **Simon P. Worden** [9], **Iaroslav Iakubivskyi** [1,10], **Mihkel Pajusalu** [10], **Laila Kaasik** [10] and **The Venus Life Finder Mission Team** [†]

1   Department of Earth, Atmospheric and Planetary Sciences, Massachusetts Institute of Technology, 77 Massachusetts Avenue, Cambridge, MA 02139, USA; jjpetkow@mit.edu (J.J.P.); rachna.agrawal.04@gmail.com (R.A.); iaroslav@mit.edu (I.I.)
2   Department of Physics, Massachusetts Institute of Technology, 77 Massachusetts Avenue, Cambridge, MA 02139, USA
3   Department of Aeronautics and Astronautics, Massachusetts Institute of Technology, 77 Massachusetts Avenue, Cambridge, MA 02139, USA
4   School of Aerospace Engineering and School of Earth and Atmospheric Sciences, Georgia Institute of Technology, Atlanta, GA 30332, USA; cecarr@gatech.edu
5   Spacefaring Technologies Pvt. Ltd., 15F, No. 14, Bhattrahalli Old Madras Road, KR Puram, Bangalore 560049, India; saragjs@gmail.com
6   School of Aeronautics and Astronautics, Purdue University, 701 W. Stadium Ave., West Lafayette, IN 47907, USA; buchanaw@purdue.edu
7   Planetary Science Institute, 1700 East Fort Lowell, Suite 106, Tucson, AZ 85719, USA; grinspoon@psi.edu
8   Fluid-Screen, Inc., 100 Cummings Center, Suite 243-C, Beverly, MA 01915, USA; monika.weber@fluid-screen.com
9   Breakthrough Prize Foundation, NASA Research Park, Building 18, P.O. Box 1, Moffett Field, CA 94035, USA; klupar@breakthrough-initiatives.org (P.K.); pete@breakthroughprize.org (S.P.W.)
10  Tartu Observatory, University of Tartu, 1 Observatooriumi, 61602 Tõravere, Estonia; mihkel.pajusalu@ut.ee (M.P.); laila.kaasik@ut.ee (L.K.)
*   Correspondence: seager@mit.edu
†   Venus Life Finder Mission Team. All memberships are listed in Acknowledgments.

**Abstract:** For over half a century, scientists have contemplated the potential existence of life within the clouds of Venus. Unknown chemistry leaves open the possibility that certain regions of the Venusian atmosphere are habitable. In situ atmospheric measurements with a suite of modern instruments can determine whether the cloud decks possess the characteristics needed to support life as we know it. The key habitability factors are cloud particle droplet acidity and cloud-layer water content. We envision an instrument suite to measure not only the acidity and water content of the droplets (and their variability) but additionally to confirm the presence of metals and other non-volatile elements required for life's metabolism, verify the existence of organic material, and search for biosignature gases as signs of life. We present an astrobiology-focused mission, science goals, and instruments that can be used on both a large atmospheric probe with a parachute lasting about one hour in the cloud layers (40 to 60 km) or a fixed-altitude balloon operating at about 52 km above the surface. The latter relies on four deployable mini probes to measure habitability conditions in the lower cloud region. The mission doubles as a preparation for sample return by determining whether a subset of cloud particles is non-liquid as well as characterizing the heterogeneity of the cloud particles, thereby informing sample collection and storage methods for a return journey to Earth.

## 1. Introduction and Motivation

The primary motivation for the Venus Life Finder (VLF) [1] Habitability Mission is to investigate Venusian cloud droplets with a variety of sensors to determine their ability to support life. This motivation arises under the assumption that, if microbial-type life exists in the atmosphere of Venus, it must be constrained to the temperate cloud layers [2–4], where the pressure is between 0.5 and 2 bar.

The key challenges to life, and hence properties to measure, are dryness and acidity. If life exists, it most likely resides inside the cloud particles, protected from a free atmosphere far drier than the limits of life as we know it [5,6]. Earth life (based on water) requires a water activity of at least 0.585 [7] for active cell division (for survival, cells can be completely desiccated for prolonged periods). However, measurements of atmospheric water vapor indicate significant variations in vertical $H_2O$ abundance (e.g., [8]; reviewed in [9]). If the relative abundance of water vapor in the temperate cloud layers can be as high as some measurements suggest, i.e., hundreds to a few thousand ppm, Venus may have pockets of high-enough humidity for life to thrive (reviewed in [9]). A gas sensor can measure atmospheric water vapor in the cloud layers to confirm or refute the presence of humid pockets. The water content inside the droplets has been inferred to be up to 30% by weight (reviewed in [10]) and could be measured by a conductivity sensor; however, water is expected to be tightly bonded to sulfuric acid.

Inside the cloud droplets, life would be protected from a fatal net loss of liquid to the atmosphere—an unavoidable problem for any free-floating microbial life forms. The composition of Venus cloud particles, however, is concentrated sulfuric acid ($H_2SO_4$), a substance fatal to most of Earth life's building blocks [11]. Therefore, one of the main goals of the VLF Habitability Mission is to measure the acidity of these particles. This investigation (i.e., determine what fraction of the particles is neutralized and to what degree) is central to Venusian astrobiology as the acidity of the cloud droplets constrains possible survival strategies of life in the Venusian clouds. The unknown composition and suspected asphericity of the Mode 3 cloud particles [12] leave room for some of the Mode 3 particles to be less acidic than pure concentrated sulfuric acid (a liquid, hence generating spherical particles). New work by [13] proposes that a subset of cloud particles may be far less acidic than previously thought, closer to a pH of 0 or 1—levels where Earth extremophiles are known to live (see also [14,15]). However, while the hypothesized neutralization of at least a fraction of the cloud particles makes them more habitable from the acidity point of view, it does not mitigate the extremely low water activity of the cloud habitat.

In addition to measuring acidity and humidity, the VLF Habitability Mission aims to identify an environment further supportive of life by searching for the presence of metals and organic compounds as well as signs of life from biosignature gases. All of these measurements aim to constrain the habitability indicators of the clouds. We note that in astrobiology the habitable environment is usually defined as an environment that contains abundant liquid water and that generally resembles habitats that are colonized by Earth-like life. However, we have to be aware that habitability is not a constant, binary characteristic of the planetary environment but rather a frontier to be explored. Many factors contribute to habitability and the VLF Habitability mission aims to characterize the Venusian cloud environment and assess how much the habitability indicators, such as water activity, cloud droplet acidity, or the availability of metals deviate from values found on Earth. Even if VLF Habitability mission measurements do not fall within the "Earth's habitability parameters", constraining these parameters would greatly expand the understanding of the chemistry of the Venusian cloud environment and constrain the feasibility of life's adaption to this environment.

## 2. Science Goals and Objectives

The overall mission science goals are: to determine habitability of the Venus cloud droplets; to search for signs of life; and to assess the cloud particle properties in order to inform capture, storage, and transfer technologies for an atmosphere sample return mission. In this section, we describe the science objectives of the mission. Table 1 connects the science goals and objectives to the instruments described in Section 3. Table 2 discusses scientific hypotheses and possible outcomes behind each science goal.

**Table 1.** Venus Habitability Mission science goals, objectives, and instruments. TLS = Tunable Laser Spectrometer, MoOSA = molybdenum oxide sensor array, TOPS = Tartu Observatory pH sensor, MEMS = micro-electro-mechanical systems, MEMS-A = MEMS aerosol elemental analyzer, MEMS-G = MEMS gas molecule analyzer, AFN = autofluorescence nephelometer. See Section 2 for a description of science goals and objectives, and Section 3 for a description of the instruments.

| | Goals | Science Objectives | Instruments |
|---|---|---|---|
| **Habitability** | **1. Measure Habitability Indicators** | 1.1 Determine the amount of water in the cloud layers | TLS and Conductivity Sensor |
| | | 1.2 Determine the pH of single cloud particles | MoOSA and TOPS acidity sensors |
| | | 1.3 Determine and identify metals and other non-volatile elements in the cloud particles | MEMS-A |
| | | 1.4 Measure the temperature, pressure, and windspeed as a function of altitude | Temperature and pressure sensor, anemometer |
| **Biosignatures** | **2. Search for Evidence of Life in the Venusian Clouds** | 2.1 Search for signs of life via gas detection | TLS and MEMS-G |
| | | 2.2 Detect organic material within the cloud particles | AFN |
| | | 2.3 Identify organic material within the cloud particles | AFN, MEMS-A |
| **Sample Return** | **3. Characterize Cloud Particles in Preparation for Sample Return** | 3.1 Determine if the cloud particles are liquid or solid | AFN |
| | | 3.2 Determine if the cloud particles are homogeneous | AFN, acidity sensors, MEMS-A |

**Table 2.** Venus Habitability Mission science outcomes.

| | Goals | Science Objectives | Hypothesis | Science Outcomes |
|---|---|---|---|---|
| **Habitability** | **1. Measure Habitability Indicators** | 1.1 Determine the amount of water in the cloud layers | The amount of water in the clouds is not uniform and is locally variable. | _Detection of anomalously high abundance values_: The amount of water in the clouds is not uniform and is locally variable. _No anomalously high values detected_: Reconciles the previous measurements and upper limits. |
| | | 1.2 Determine the pH of single cloud particles | The acidity of cloud particles is variable and not all cloud particles are composed only of concentrated sulfuric acid. | _Detection of variable acidity of cloud particles_: The altitude profile of cloud acidity tests the validity of atmospheric and cloud models and model implications for the habitability of the clouds. _Acidity of cloud particles is uniform and consistent with concentrated sulfuric acid_: Puts clear constraints on the chemical processes in the atmosphere; confirms, for the first time by direct measurement, that the clouds are uniformly made of concentrated sulfuric acid particles. |
| Biosignatures | | 1.3 Determine and identify metals and other nonvolatile elements in the cloud particles | Cloud particles contain dissolved metal ions (e.g., Fe) and other ions of nonvolatile elements (e.g., P). | *Metal ions detected*: The composition of the cloud particles is chemically complex; Suggests efficient exchange of material between the surface (the presumed source of the non-volatile elements) and the clouds.<br>*No metal ions detected*: The material exchange between the surface (the presumed source of the non-volatile elements) and the clouds is not efficient limiting the habitability of the clouds. |
| | | 1.4 Measure the temperature, pressure, and windspeed as a function of altitude | No specific hypothesis; more direct measurements are needed to understand the atmospheric dynamics. | Atmospheric dynamics can be compared to those predicted by current general circulation models of Venus (e.g., [16]) and are important to inform operations for future missions such as balloons or aerobots as well as aerosol sampling in support of in-situ analysis and/or sample return. |
| | **2. Search for Evidence of Life in the Venusian Clouds** | 2.1 Search for signs of life via gas detection | Gases listed in Table 3 are signs of chemical disequilibria in the clouds that could be associated with life. | *Disequilibrium gases (Table 3) detected*: The abundance vs altitude profile constraints sources of gases and tests the validity of the atmospheric chemistry models and their implications.<br>*Non-detection of disequilibrium gases (Table 3)*: Reconciles the upper limits provided by the remote observations with the tentative in situ detections; Puts clear constraints on the chemical processes in the atmosphere. |
| | | 2.2 Detect organic material within the cloud particles | Clouds of Venus are not a chemically sterile environment and contain organic molecules. | *Organics detected*: The prospect for the habitability of Venus' clouds increases as all life requires organic chemistry.<br>*No organics detected*: The prospects of the clouds of Venus as a habitable environment diminish as we assume that all life, no matter its chemical makeup, requires organic chemistry. |
| | | 2.3 Identify organic material within the cloud particles | Cloud particles contain complex organic molecules that could be precursors to life or even be signs of life itself. | *Complex and diverse organics identified*: Potential for life in the cloud particles increases with the diversity and complexity of detected organics.<br>*Only simple and uniform organics identified*: Abiotic processes are most likely responsible for organics formation.<br>*No organics identified*: The prospects of the clouds of Venus as a habitable environment diminish as we assume that all life, no matter its chemical makeup, requires organic chemistry. |
| Sample Return | **3. Characterize Cloud Particles in Preparation for Sample Return** | 3.1 Determine if the cloud particles are liquid or solid | Clouds are not homogenous and are composed of liquid concentrated sulfuric acid droplets and solid salt particles. | *Solid Mode 3 particles composed of salts detected*: Confirms the existence of the Mode 3 particles; the salt composition puts clear constraints on the chemical processes in the cloud droplets and the atmosphere; confirms that the clouds are not uniformly made of liquid concentrated sulfuric acid particles.<br>*No solid particles detected*: Supports the model that the clouds of Venus are made of liquid droplets of concentrated sulfuric acid. |
| | | 3.2 Determine if the cloud particles are homogeneous | Cloud particles are not homogenous in terms of shape, size, acidity and chemical composition. | *Cloud particles' chemical composition and acidity vary*: Cloud particles are non-homogenous. Informs the design of the future particle capture and storage technology for the atmospheric sample return mission.<br>*Cloud particles' chemical composition and acidity are uniform*: Supports the model that the clouds of Venus are made of liquid droplets of concentrated sulfuric acid. Informs the design of the future particle capture and storage technology for the atmospheric sample return mission. |

### 2.1. Goal 1: Measure Habitability Indicators

#### 2.1.1. Objective 1.1: Determine the Amount of Water Vapor in the Cloud Layers

Water ($H_2O$) is of interest because the prevailing view is that Venus clouds are uniformly an incredibly dry environment. Water vapor in the clouds, however, cannot be globally homogeneous based on measured variations in vertical $H_2O$ abundance [13]. Some measurements even suggest localized "pockets" of humidity of hundreds to even a few thousand ppm (reviewed in [9]). If, however, those localized in situ measurements are erroneous, and if life on Venus is based on water, then the ultra-dry and hyper-acidic cloud conditions are too extreme for life as we know it [5,6]. To survive, life would require evolutionary adaptations never developed by life on Earth.

In addition to measuring water vapor, it is necessary to determine the water content of the cloud particles. The motivation to measure the water content of the cloud particles stems from the in situ measurements of water from the 1970s and 1980s. Soviet VeGa and Venera probes and NASA Pioneer Venus give abundances of 200–2000 ppm in the middle/lower clouds (48–58 km) and 5000 ppm just below the clouds (41.7 km) [9]. If correct, they could suggest that such localized anomalously high water abundances are relatively common, as both the Pioneer Venus as well as Venera probes did detect them, although ascribing probability to such observations is difficult. These values are considerably higher than the global average of around 30 ppm, obtained below the clouds, from the spectrometric measurements [17] (the average value is approximately ten times lower above the clouds [18–21]). Mogul et al. 2021 [15] suggest the high $H_2O$ abundance values obtained by the in situ measurements are due to water from the cloud particles themselves. If such an interpretation is true, it means the cloud particles have more water than our current understanding allows. Ultimately such discrepancies can only be resolved by new in situ measurements of water abundance at multiple locations in the atmosphere.

#### 2.1.2. Objective 1.2: Determine the Acidity of Single Cloud Particles

According to current consensus, Venusian cloud particles are predominantly liquid concentrated sulfuric acid droplets [10]. The droplets would then be extremely acidic, billions of times more so than the most acidic environments supporting life on Earth. No Earth life as we know it can survive in droplets of such high acidity, although life with protective shells might survive. (Shells could be composed of material known to survive in sulfuric acid, e.g., certain lipids [22], sulfur [23], or graphite).

Low single-droplet acidity for a large number of cloud particles would be a key indicator of habitable conditions. There is a strong motivation to test the longstanding high-acidity consensus by investigating whether the acidity of some cloud droplets is more favorable for life as we know it than that of concentrated sulfuric acid.

New work proposes that mineral dust carried from the surface by updrafts or locally-produced $NH_3$ could dissolve in sulfuric acid droplets, buffering them to habitable levels [13–15]. If the droplets have a pH of 0 or 1, their acidity is consistent with the most acidic environments where life is found on Earth.

If droplets are less acidic than previously thought, we need to know what fraction of particles is neutralized and to what degree. The question of homogeneity is important for a future sample return mission: are all particles the same or are a fraction different with properties that make them potentially habitable? The answer will inform sample capture and storage technologies and volume, and will help establish the altitudes of interest for sample capture.

#### 2.1.3. Objective 1.3: Detect and Identify Metals and Other Non-Volatile Elements in the Cloud Particles

Life as we know it requires metals and other non-volatile species (e.g., phosphate). Detection of such species as components of cloud particles raises the potential for habitability of the clouds. Possible VeGa and Venera detections of Fe and P tentatively suggest a heterogeneous composition of dissolved ions and chemicals in cloud particles [24,25]. (While the balloon implementation of our proposed mission will not sample the ~60 km altitude where the "unknown UV absorber" [10] absorbs the most, it is worth noting that the UV absorber might include metals such as Fe in $FeCl_3$ [26]).

Measuring the elemental composition of the aerosols as a function of the altitude and searching for organic material within the particles will inform the priority altitudes for sample capture for the atmosphere sample return mission.

### 2.1.4. Objective 1.4: Measure the Temperature, Pressure, and Windspeed as a Function of Altitude

The measurements of temperature-pressure profiles and wind speed in the clouds of Venus are not directly astrobiological in nature but provide essential context and constraints on relevant processes. Transient planet gravity waves are encoded in the temperature pressure profiles (e.g., [27]), and measuring them helps substantiate the concept of moving material, including hypothetical spores, up from lower atmosphere layers [5]. Vertical air movements are a general habitability requirement for any aerial biosphere. As cloud particles collide with one another, they grow until they become massive enough to gravitationally settle down out of the clouds and must be transported back up to the temperate regions [5].

### 2.2. Goal 2: Search for Evidence of Life in the Venusian Clouds

### 2.2.1. Objective 2.1: Search for Signs of Life via Gas Detection

Several gases are of crucial interest for habitability or even as signs of life. The Habitability Mission aims to detect reduced and anomalous gas molecules as a sign of disequilibrium chemistry and as potential biosignatures (Table 3).

**Table 3.** Gases of interest. TLS = tunable laser spectrometer, MEMS-G = MEMS gas molecule analyzer. See also [9] for the detailed discussion of the Venusian atmospheric anomalies, including anomalous gas detections.

| Gas | Motivation | Scientific Outcomes | Instrument |
|---|---|---|---|
| $O_2$ | Potential sign of life; prior in situ detection | _Detection_: The abundance vs altitude profile constraints the source of $O_2$ and tests the validity of the models and their implications. <br> _Non-detection_: Reconciles the upper limits provided by the remote observations with the tentative in situ detections; Puts clear constraints on the chemical processes in the atmosphere. | MEMS-G, TLS |
| $SO_x$ | Variable profile indicative of unknown cloud particle chemistry | _Detection_: The abundance vs altitude profile constraints the source of $SO_2$ and other $SO_x$ gases and tests the validity of the models and their implications; Puts clear constraints on the chemical processes in the cloud droplets and the atmosphere. | MEMS-G |
| $NO_x$ | Important component of the nitrogen cycle; prior tentative detection | _Detection_: The abundance vs altitude profile constraints the source of $NO_x$ and tests the validity of the models and their implications. <br> _Non-detection_: Puts clear constraints on the chemical processes in the cloud droplets and the atmosphere, including on the presence and intensity of lightning strikes. | MEMS-G |
| $H_2O$ | Variable profile indicative of unknown cloud particle chemistry, including some anomalously high values | _Detection of anomalously high abundance values_: Confirmation that the amount of water in the clouds is not uniform and is locally variable. <br> _No anomalously high values detected_: Reconciles the values and upper limits provided by the remote and in situ spectroscopic observations with the tentative in situ detections. | TLS |
| NH$_3$ | Potential sign of life; habitability indicator; potential neutralizing agent for cloud droplets; prior tentative detection | _Detection_: The abundance vs altitude profile constraints the source of NH$_3$ and tests the validity of the models and their implications.<br>_Non-detection_: The NO$_x$ species (if confirmed) could not be the result of oxidation of NH$_3$; Reconciles the upper limits provided by the remote observations with the tentative in situ detections; Puts clear constraints on the chemistry of the cloud droplets and on the chemical processes in the atmosphere. | TLS or MEMS-G |
| PH$_3$ | Potential sign of life; prior tentative detection | _Detection_: The abundance vs altitude profile constraints the source of PH$_3$ and tests the validity of the models and their implications.<br>_Non-detection_: Reconciles the remote observations, including the upper limits, with the tentative in situ detections; Puts clear constraints on the chemical processes in the atmosphere, including the availability of volatile P species. | TLS |
| CH$_4$ | Potential sign of life; prior anomalous detection | _Detection_: The abundance vs altitude profile constraints the source of CH$_4$ and tests the validity of the models and their implications; Provides a potential source for organic chemistry in the clouds.<br>_Non-detection_: Reconciles the upper limits provided by the remote observations with the tentative in situ detections. | TLS or MEMS-G |

### 2.2.2. Objectives 2.2 and 2.3: Detect and Identify or Constrain Organic Material within the Cloud Particles

The presence of organic chemicals is crucial for any life to exist. Therefore, any detection and identification of organic compounds in the atmosphere of Venus would bolster support of the habitability of the clouds. The search for organic molecules in Venus cloud particles has not yet been attempted.

Autofluorescence is a way to detect (and potentially identify) organic compounds. When subjected to UV light, organic molecules with conjugated double bonds (e.g., aromatic molecules that have delocalized electrons in the rings) generally yield stronger fluorescent signals compared to other molecules. While inorganic salts and other materials can also fluoresce, the signal is expected to be much lower than for molecules that contain organic aromatic rings, and likely too low for detection.

### 2.3. Goal 3: Characterize Cloud Particles in Preparation for Sample Return

#### 2.3.1. Objective 3.1: Determine If the Cloud Particles Are Liquid or Solid

For a future atmospheric sample return mission, we need to understand if any small subset of the cloud particles are solid, as the ideal capture methods and storage substrates may differ for liquid and solid particles. The majority of the Venus cloud particles are liquid sulfuric acid.

The Venus Mode 3 cloud particles have a component of unknown composition and are aspherical [12]. Given that an aspherical particle cannot be purely liquid, the possible non-liquid nature of Mode 3 particles is of significant interest. Rimmer et al.'s theory [13] suggests that some Venus cloud particles are solid slurries with a pH of ~1. A confirmation of the shape of the Mode 3 particles, along with measurements to constrain the composition, will help determine if the Mode 3 particles have habitable properties.

#### 2.3.2. Objective 3.2: Determine If the Venus Cloud Particles Are Homogeneous

Our current understanding, based on extremely limited in situ data, is that the clouds on Venus are homogeneous but separated into three particle Modes whose relative fraction differs with altitude. Any further degree of inhomogeneity will inform sample return, namely, the required sample volume and storage technology. In other words, will a small batch sample be sufficient, or should we aim to collect many of the "one-in-a-million" particles that might host life?

## 3. Instrument Summary

The instrument suite is designed to sample the atmosphere to meet the science goals and objectives described in Section 2 and Table 1. Instruments outside of a pressure vessel must be built to withstand the acidic conditions of the clouds of Venus.

### 3.1. Single Particle pH Meter

A single-particle pH meter must be custom-developed for the Venus atmosphere. Currently, no instrument exists, that would suit the mission's requirements, that can measure pH in negative numbers accurately or across the acidity range anticipated on Venus (which we aim to measure in full). If some cloud particles are partially neutralized, they may have a pH of −1 or higher. However, if the droplets are composed of liquid concentrated sulfuric acid, we expect to measure Hammett acidity ($H_0$) of −11 or lower. Our main goal is to distinguish between droplets with an acidity well below zero and those with pH greater than zero. Whether the pH of a semi-solid particle can be measured as accurately as that of a liquid particle is still in question. We consider two different pH meter concepts.

The Tartu Observatory pH Sensor (TOPS) will measure the acidity of single Venusian cloud droplets by the established method of fluorescence spectroscopy, which is widely used for pH measurements. For a detailed description of the TOPS acidity sensor, please see [28].

In brief, the general mechanism for the single particle acidity sensor is using a dye-sensitized sensor plate and illuminating it with various wavelengths of light. After cloud particles hit the sensor plate, different spots will fluoresce with different intensities at a given excitation wavelength, allowing for the measurement of single-particle pH.

Fluorescein is a candidate fluorophore for the TOPS sensor due to its high fluorescence intensity [29] and stability over a vast range of sulfuric acid concentrations, from dilute to highly concentrated solutions (e.g., over 10 M sulfuric acid) [30,31]. Under strongly acidic conditions, fluorescein is in its cationic form. Since emission responses of fluorescein forms differ drastically, fluorescein is a good candidate as a fluorescence dye for the Venusian single-particle pH meter [29,31].

The TOPS sensor is a one-use instrument that cannot be reused. The sensor film has to be protected during the voyage to Venus and is exposed to the atmosphere only while measuring the acidity of the single droplets hitting the sensor plate. When the sensor plate is saturated with impacting droplets, the sensor plate becomes unusable and would be powered off. For the Venus habitability mission, two or three sensors could be considered that would measure the acidity of single droplets at different altitudes. Alternatively, the system could be equipped with an external circular array of sensor plates that can be exposed at various altitudes using the same instrument. Studies on the operation time and the number of particles that a single sensor plate can detect are currently underway [28].

The molybdenum oxide sensor array (MoOSA) pH sensor can categorize individual cloud particle pH values using an optical method that is mechanically and chemically robust and can measure minute sample volumes.

MoOSA will categorize individual droplets deposited on its surface into one of three acidity categories: pH < 0; pH 0–1; pH > 1. MoOSA will return counts in each pH category, corresponding to individual depositions on an array of plasmonic $MoO_3$-coated micro-ring resonators (MRRs) [32]. MoOSA sensor can be refreshed by applying a small voltage to the sensitive film, allowing it to be reused throughout the mission's duration.

Both TOPS and MoOSA consist of a sensor plate respectively that is exposed to the clouds of Venus. The rest of the sensor systems are protected from the concentrated sulfuric acid environment. Both sensors are lightweight with TOPS weighing about 350 g and MoOSA about 200 g, and both require a power of about 6 W (see Table 4).

**Table 4.** Scientific instruments for the Venus Habitability Mission.

| Instrument | Mass ** (kg) | Volume (cm³) | Average Power (W) | * Data Vol. per Meas. (kB) | TRL |
|---|---|---|---|---|---|
| Mini tunable laser spectrometer (TLS) | 4.60 | 240 | 14.0 | 1000 | 4 |
| MEMS aerosol analyzer (MEMS-A) | 0.34 | 400 | 1.0 | 27 | 4 |
| Autofluorescing nephelometer (AFN) | 0.80 | 100 | 40.0 | 120 | 3 |
| Tartu Observatory pH Sensor (TOPS) | 0.35 | 844 | 2.0 | 1 | 2 |
| Imaging unit (IU) | 0.15 | 250 | 0.5 | 100 | 4 |
| Weather Instruments Suite (WIS) | 0.10 | 98 | 1.0 | 0.05 | 4 |
| **Total Gondola Subsystem Mass** | **6.34** | **1932** | **58.5** | **1248** | |

* The combined data volume of all the instruments per measurement is about 1.3 MB. The data volume transmitted to Earth depends on whether the communication is direct-to-Earth or through an orbiter relay. For a balloon mission, the data volume can vary from 25 MB to 250 MB for both options respectively (see detailed discussion in [33]). On-board data processing will be required to consolidate the measurements and select "interesting" data sets for transmission so that more information can be transferred in less data volume. ** Mass includes supporting sample collection and handling systems. Table from [22].

## 3.2. Autofluorescence Nephelometer

The autofluorescence nephelometer (AFN) will constrain the cloud particle size distribution and composition. The AFN will shine a UV laser on the sample to induce autofluorescence in any organic material inside cloud particles. Autofluorescence will indicate the presence of organic materials in the cloud droplets because organic compounds with delocalized electrons in ring structures yield stronger fluorescent signals when subjected to UV light than other molecules. The cloud particle samples need no handling and can be accessed by shining the UV laser through a pressure vessel window. Another option is to allow the airstream to pass through a narrow tube with an optical window for laser illumination. While more complex, the geometry with an ingested sample would permit a shorter working distance and as a result, greater sensitivity.

In addition to fluorescence, a measurement of the intensity of the laser light and polarization backscattered off of the particles will be used to constrain the composition and shape of the particles. Suppose we can confirm past measurements that indicate some cloud particles are aspherical (i.e., not liquid). In that case, it affirms that some cloud particles are not pure concentrated sulfuric acid and would represent a currently unknown composition.

The AFN is being built by Droplet Measurement Technologies (DMT) for the Rocket Lab mission to Venus [34] and has considerable heritage via DMT's commercial products that fly on the outside of aircraft. The backscatter cloud probe (BCP) [35] has over 60,000 h of data acquired on a global basis [36]. The BCP was subsequently upgraded to include the detection of polarization, and the AFN takes this further by the addition of a fluorescence detection channel. The AFN is a power-hungry instrument with an operating power requirement of 40 W (dominated by temperature stabilization), but is lightweight at about 800 g. To go beyond the detection of organic compounds towards identification, we may consider adding an additional laser and detector and even a Raman spectroscopy component to the AFN. For a detailed description of the AFN development see the dedicated article in the same special issue [37].

### 3.3. MEMS Aerosol Elemental Analyzer

A Micro-Electro-Mechanical Systems (MEMS) Aerosol Elemental Analyzer (MEMS-A) can identify ions dissolved in cloud droplets. MEMS devices are lightweight and low power, thus attractive to our mission. A MEMS-A instrument must be tailored to specific chemical species. Our main targets are oxidized P species and metals, including Fe, Mg, Ca, Mn, Cu, Na, and K.

An example instrument is the MEMS ion-gas micro-spectrometer with optical signal detection for aerosol analysis (MEMS-A) [38,39]. The MEMS-A is ~340 g with a power requirement of 4 W. The sample inlet system consists of a capillary channel with a wide-ened inlet. Its microfluidic channel and a mechanical MEMS micropump allow for the flow of the sample to the sample ionizer, which ionizes the sample and induces light emission. The emission from the ionized sample is detected by the MEMS spectrometer, which allows for qualitative (based on the wavelength of emission peaks) and quantitative (based on the measured intensity of emission peaks) identification of chemical species in the sample mixture. The result of a single measurement is a VIS/NIR emission spectrum.

The MEMS-A device can detect ppm to ppb abundance of specific metals and other ions dissolved in the sample (the detection limit depends on the particular metal ion). One measurement of sample composition requires at least 34 droplets with a mean diameter of 1 μm. Measurements for acidic aerosols below ~100 hPa (0.1 bar) cannot be performed. Since the cloud decks span the atmospheric pressure range of approximately 0.5 to 2 bar, the MEMS pressure limitations are not a serious issue.

An alternative to MEMS-A is an X-ray fluorescence spectrometer (XRF). X-ray fluorescence is a non-destructive analytical technique used to determine the elemental composition of the sample. X-ray fluorescence occurs when a fluorescent (or secondary) X-ray is emitted from a sample that is being excited by, e.g., a primary X-ray source. Because X-ray fluorescence is unique to the elemental composition of the sample (the emitted X-ray photon's energy increases with atomic number), XRF is an excellent technology for qualitative and quantitative analysis of the material composition. Soviet VeGa and Venera probes used XRF spectroscopy to analyze the elemental composition of the cloud particles retained on the filter and detected non-volatile elements like iron or phosphorus [24,25]. A similar, more modern and miniaturized XRF device could be used for elemental analysis of the cloud particles for the VLF habitability mission.

### 3.4. Mini Tunable Laser Spectrometer

The tunable laser spectrometer is a mature in situ gas detection instrument used in a variety of commercial applications, including medical sensing, industrial sensing, and Earth science [40,41]. The TLS has seen tremendous success on the Mars Curiosity Rover [42,43].

The TLS infrared laser absorption spectrometer has a small gas cell with highly reflective surfaces so that a laser can bounce back and forth thousands of times to construct an effective path length of up to 10 km. By enabling such a long path length at gas cell pressure below ~100 mbar, the TLS can measure ppb-level gas abundances at very high spectral resolution ($\lambda/\Delta\lambda$ = 10 million) which enables the detection of individual rovibrational lines). The unique advantages of the TLS are its technological maturity and completely unambiguous detection of a given molecule.

The mini tunable laser spectrometer is a miniaturized version of the TLS under development at JPL (Chris Webster, PI). While the mini TLS has not yet been flown in a space mission, it traces extensive heritage. The mini TLS considered for the mission has four channels, each of which can cover a wavelength range dedicated to a gas of interest. Additional molecules also might fall within a specially chosen wavelength range. Because the channels have a narrow wavelength range (a spectral width of about 6 cm$^{-1}$), the target gases of interest must be selected in advance. Mini TLS is the heaviest instrument in the suite, with a mass of about 4.6 kg and a power requirement of about 48 W if all channels operate simultaneously (Table 4). The sample inlet for the instrument will have a filter to prevent sulfuric acid from entering the gas cell. A pumping system will transfer the sample to the gas cell.

### 3.5. MEMS Gas Molecule Analyzer

A MEMS gas molecule analyzer (MEMS-G) must be tailored towards the detection of a single specific gas, and placed on a deployable mini probe. A MEMS-G will need to be able to measure $O_2$ to low-ppm levels and $NH_3$ and $CH_4$ down to ppb levels.

One example of a MEMS-G is a device similar in design to the MEMS-A, the MEMS ion-gas micro-spectrometer with optical signal detection for gas mixture analysis (MEMS-G) [38,39], which can detect trace gases in the atmosphere to a limit of detection of 0.1 to 0.01 ppm (depending on the expected gas composition and the atmospheric pressure). MEMS-G has a similar mass, power, and inlet-transfer system to the MEMS-A device. The gas sample is ionized inside the MEMS electron-impact ionizer to induce light emission. Peaks of emission, detected by a MEMS spectrometer, enable the identification of gases. The MEMS devices have pressure-dependent sampling times ranging from 1 min for low pressures up to a maximum of 10 min (including sample acquisition). MEMS-G was originally developed to identify the isotopic composition of Martian $CH_4$.

Another example of a MEMS-G-type instrument is the Tartu Observatory $O_2$ sensor, described in [44].

### 3.6. Weather Instrument Suite

The Weather Instruments Suite (WIS) will measure temperature, pressure, and wind speed. An example of a commercially available combined temperature and pressure sensor is the Paine 211-55-010 Series Pressure Transducer from Emerson's Rosemount Inc., which is rated to operate at up to 344 bars and 316 °C [45]. This instrument is suited to cover temperature measurements from 80 km to 20 km and pressure measurements from 70 km to 20 km, thereby encompassing the entire cloud deck and stagnant haze layer below the clouds. An example anemometer is a miniaturized 3D wind sensor being developed for the Martian atmosphere [46]. The WIS instruments are extremely lightweight with a total mass of less than 100 g and require total power of less than 1 W. All the WIS sensors are exposed to the atmosphere for direct measurements.

### 3.7. Conductivity Sensor

To complement measurements of atmospheric water vapor it is also important to determine the water content of the cloud particles. One instrument under consideration is a modified conductivity sensor that draws inspiration from the TOPS single particle acidity sensor. A conductivity sensor can measure the ability of a solution to conduct an electrical current, and therefore provides a measure of the solution's dissolved ion composition. Such sensors are routinely used in pure water monitoring, chemical processing, and the pharmaceutical industry. However, more work is needed to assess the required adaptations of such a sensor to the Venusian concentrated sulfuric acid environment (i.e., sulfuric acid concentration and potential ion composition).

The MEMS devices and mini TLS require sample collection, which has its own challenges in the Venus environment. The inlet must be able to prevent clogging and allow smooth flow of the acquired sample. Clogging of sample inlets has been a serious challenge in previous in situ missions to Venus. In the Venera-9 mission, micro-pore intakes used in the mass spectrometer were clogged by the cloud particles. The particles evaporated at lower altitudes due to high temperatures. In the subsequent Venera-11 mission, intakes with large circumferences were used to prevent clogging. The Pioneer Venus

LNMS inlets also suffered from clogging. The planned DAVINCI mission attempts to solve the inlet clogging problem by "implementing heated inlet tubes to vaporize trapped droplets or filters to capture cloud particles large enough to cause potential clogs. DAVINCI mission also supports a second dedicated inlet for sampling atmosphere below the sulfuric acid cloud and haze layers" [47].

The MEMS and TLS devices will require similar innovations to prevent clogging by the cloud particles.

## 4. Mission Implementation Concepts

The Venus Habitability Mission instrument suite can be hosted by a variety of mission architectures. Companion papers to this article describe the mission architectures [33] and aerial platforms [48], which are summarized here.

### 4.1. Fixed Altitude Balloon Concept

A fixed altitude balloon with mini probes is the baseline concept for the Venus Habitability Mission. The design of the mini probes is still being traded but can build on concepts such as a LEVL probe [49] or a LOVE-Bug [50]. A one-week balloon mission permits horizontal atmospheric sampling at an altitude of 52 km (Figure 1). The super-rotating atmosphere of Venus enables the balloon to circumnavigate the planet in four Earth days, thus resulting in sampling at various longitudes and on both day and night sides of the planet. In addition, four mini probes (Table 4) deployed from the balloon in pairs can sample the atmosphere vertically—at two different latitudes—to assess the chemical properties of the clouds across the cloud deck and below the clouds. The winds on Venus also have a North-South component which enables the balloon to travel about 12 degrees of latitude in one circumnavigation in a near-equatorial region. The balloon altitude also cycles naturally within 1–2 km of the design altitude due to thermal variations (VeGA balloons' altitude varied from 51 to 54 km). The mini probes (Table 5) can measure the vertical profiles of selected gases, single droplet acidity, and search for the presence of metals, complementing the measurements performed by the balloon gondola instruments. Each mini probe will further include a UHF radio transmitter, batteries, and a parachute. This fixed-altitude balloon and mini probes discussed here are described in [22,33]. Aerial platform considerations are described in [48]. The mission duration can potentially be extended to up to a month, depending on power availability, permeation of the lifting gas, and survivability of the balloon and associated payloads.

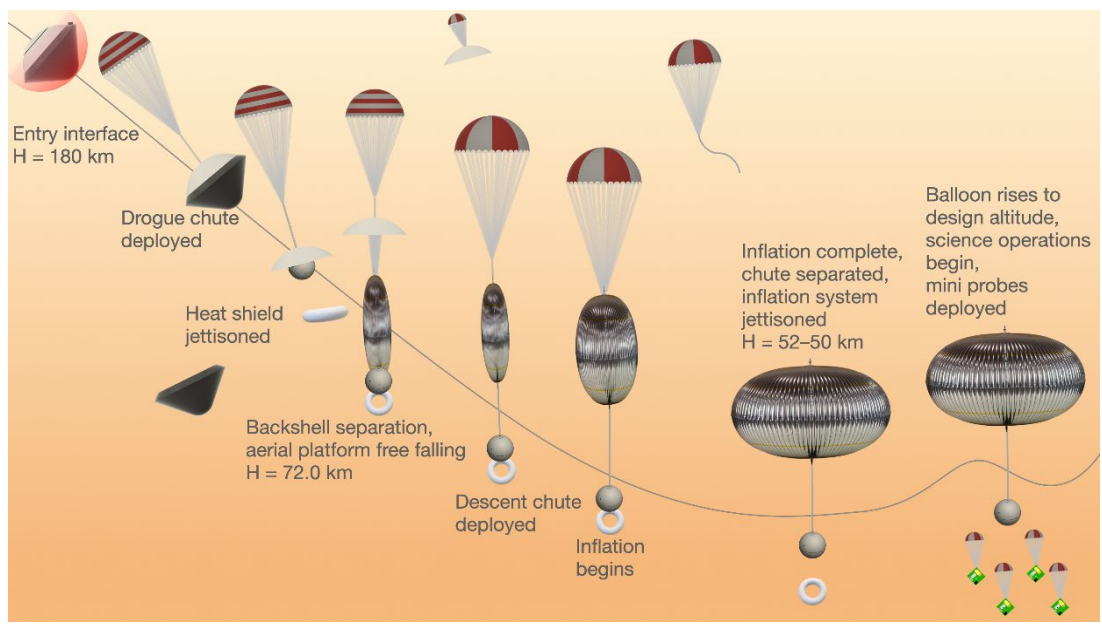

**Figure 1.** Entry, descent, and balloon deployment and operation sequence for a fixed-altitude balloon and mini probes. Not to scale. Figure modified from [48].

**Table 5.** Scientific instruments considered as payloads for the mini probes.

| Instrument | Mass (kg) | Volume (cm³) | Average Power (W) | Data Vol. per Meas. (kB) | * TRL |
|---|---|---|---|---|---|
| MEMS gas analyzer (MEMS-G) ** | 0.34 | 400 | 0.8 | 27 | 4 |
| MEMS aerosol analyzer (MEMS-A) | 0.34 | 400 | 1.0 | 27 | 4 |
| Tartu Observatory pH Sensor (TOPS) | 0.35 | 844 | 2.0 | 1 | 2 |
| MoOSA pH sensor (MoOSA) ** | 0.20 | 10 | 2.0 | 1 | 2 |
| Weather Instruments Suite (WIS)—one in each mini probe | 0.10 | 98 | 1.0 | 0.05 | 4 |
| **Total Mini Probe Instrument Mass** | **1.63** | **2046** | **9.80** | **56.20** | |

* Mass includes supporting sample collection and handling systems. ** Due to their relatively small mass and size MEMS-G and MoOSA instruments are particularly suited as payload for the mini probes. Table from [22].

### 4.2. Variable Altitude Balloon Concept

A variable altitude balloon is a second mission implementation concept for the Venus Habitability Mission. Vertical excursions enable the instrument suite to sample at different altitudes (Figure 2). The variable altitude balloon mission concept is described in [22,33,48].

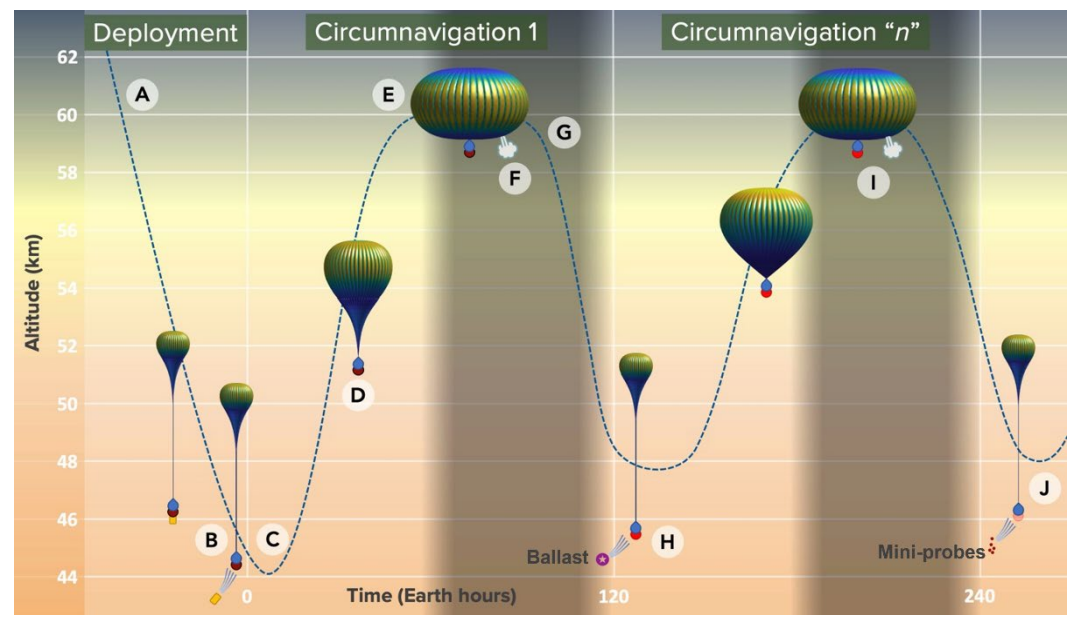

**Figure 2.** Balloon deployment and operation sequence for a variable altitude balloon, an alternative to the balloon mission shown in Figure 1. Not to scale. Figure from [33]. See main text of [33] for further details and the description of steps **A**-**J**.

### 4.3. VLF Habitability Parachute Probe

A large probe with a parachute (and no balloon) on a single vertical descent is adequate to carry out the Venus Habitability Mission science goals. A ~6 m parachute would enable a 60 kg probe to last about one hour in the cloud layers (40–60 km altitude). Two large probes carried on the same cruise vehicle for efficiency would allow sampling of different locations as well as redundancy. A large probe avoids the complexity of the balloon system and has heritage from Pioneer Venus, but unlike a balloon system has limited latitudinal spatial sampling and no technology maturation path for a sample return mission [22].

We provide a representative concept of operations for the parachute mission architecture (Figure 3). The cruise vehicle approaches Venus on a hyperbolic trajectory. The entry system separates from the cruise vehicle on an entry trajectory. We define the entry interface at 180 km and anticipate an entry speed of 11.2 km/s. Shortly after entry, a mortar deploys a pilot chute which extracts the aft shell. In turn, the aft shell extracts the main chute. A few seconds later, explosive bolts separate the probe from the forward aeroshell heat shield. Science operations begin near the top of the cloud layer and conclude at approximately 48 km altitude, when the probe transmits gathered data to the cruise vehicle as it flies by the planet.

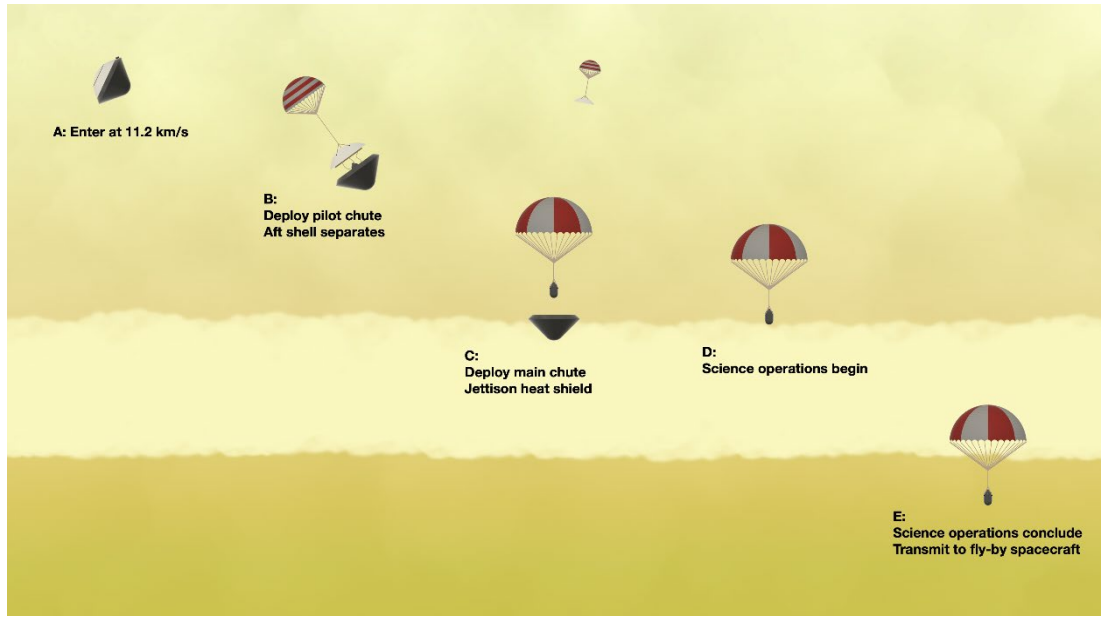

**Figure 3.** Representative concept of operations for the VLF Habitability parachute probe Habitability Mission.

The baseline probe for the VLF Habitability Mission accommodates the instrument suite presented in Table 4. The exact design will depend on the mission architecture as thermal protection considerations, energy requirements, and data rate capabilities are functions thereof. We provide a mass breakdown of the Habitability Mission probe in Table 6.

**Table 6.** Mass breakdown of the VLF Habitability Mission parachute probe concept.

| Component | CBE Mass (kg) | Contingency | MEV Mass (kg) |
| --- | --- | --- | --- |
| Structure | 2.3 | 1.3 | 3.0 |
| Science Instruments | 6.4 | 1.3 | 8.3 |
| Battery + PDS | 0.4 | 1.3 | 0.5 |
| Communication | 3.7 | 1.3 | 4.8 |
| Thermal | 1.2 | 1.3 | 1.6 |
| C&DH | 3.1 | 1.3 | 4.0 |
| **Total** | **17.1** | | **22.3** |

We present a notional design for the baseline probe in Figure 4. The probe is a capsule shape with radius of 10.3 cm and cylindrical section height of 15.2 cm. The hull wall is 2 mm thick and is lined with a thin layer of Kapton. Aft and forward beryllium shelves coated in sodium silicate accommodate the science payload and other necessary subsystems. There are 12 sealed penetrations in the pressure vessel: 3 inlets and 3 outlets for the TLS, AFN, and MEMS-A devices; 5 protrusions for the AN, TP, TOPS, MoOSA, and ANT; and a window for the IU. The total mass of the probe likely will not exceed 22.3 kg, with the current best estimate being 17.1 kg.

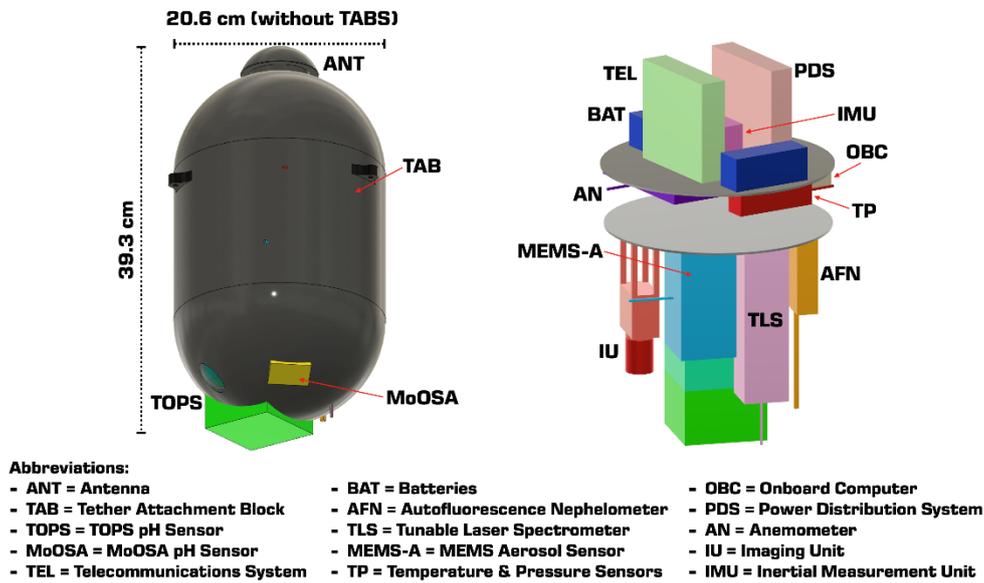

**Figure 4.** Concept design for the VLF Habitability parachute probe Habitability Mission.

In all three mission architectures, the primary payload suite is common since each instrument can make measurements at a high rate—on the order of 1 Hz. This allows for all the instruments to operate on the parachute probe as well. The difference between the architectures, in terms of science return, is the number of measurements and the spatial and temporal variation in the samples. The balloon-based missions would provide a much larger data budget and time for communication, while the parachute mission would require data compression and more sophisticated autonomy.

### 4.4. Alternative Science Payload

Detection and identification of complex organic chemicals are considered to be indicative of the presence of life (e.g., [51]). In order to identify complex molecules in the cloud particles of Venus, we need to go beyond the AFN, whose aim is to establish the presence of organic molecules without actually being able to identify them.

The instrument of major interest is a laser desorption mass spectrometer (LDMS). Laser desorption mass spectrometry can analyze almost any solid material without the need for prior pyrolysis of the sample. This includes, but is not limited to potential biochemicals (nucleobases, amino acids, lipids, metabolic molecules, etc.), salts, minerals, and other solid residues present in atmospheric particles.

While several space missions have utilized mass spectrometers, all have relied on pyrolysis to date. We choose to avoid the destructive process of pyrolysis to allow for more reliable identification of parent complex organic molecules.

Two example LDMS instruments being designed for space are *CORALS* [52] and *ORIGIN* [53,54]. These instruments can detect trace levels (~fmol) of complex organic compounds (>200 Da) from chemically diverse samples without prior knowledge of the sample chemical composition—a feature crucial for chemical characterization of the unknown environment of the clouds of Venus. The LDMS has a high vacuum chamber (with vacuum levels required at $5 \times 10^{-8}$ mbar [54]) for sample analysis, introducing a major design challenge for any sample delivery system. A third LDMS instrument is the Mars Organic Molecule Analyzer (MOMA) instrument planned for the ExoMars rover [55].

Instead of the instrument suite presented in Table 4, a VLF Habitability Mission could feature a probe accommodating only an LDMS as its science payload. We provide a mass breakdown of the LDMS descent probe in Table 7. While offering a lesser variety of data, the potential science return associated with the identification of complex molecules within the Venusian cloud layer is tremendous.

The principal design challenge that must be overcome in order to incorporate an LDMS system into a Venus descent probe mission lies in the mechanisms of sample collection and delivery to the instrument. One simple way to acquire a liquid sample is by a passive collection of liquid aerosols onto a mesh-like wire surface, using gravity to move the liquid through pipes to storage vials. The efficiency of the passive collection is relatively low but can be increased by optimizing mesh geometry and implementing cloud ionization, ensuring that electrostatic force overcomes aerodynamic limitations. We estimate that a 30 × 30 cm passive collector would be able to acquire more than ten milliliters of liquid in lower clouds per day of sampling. This number decreases to under 0.5 mL per day in upper cloud layers. One milliliter of collected liquid contains $10^9$–$10^{13}$ droplets. Clogging and contamination must be actively avoided. To provide a suitable sample to an LDMS, the liquid may need to be filtered or heated and non-sulfuric acid components concentrated before transfer to the LDMS.

**Table 7.** Mass breakdown of the probe containing the LDMS instrument.

| Component | CBE Mass (kg) | Contingency | MEV Mass (kg) |
|---|---|---|---|
| Structure | 6.6 | 1.3 | 8.5 |
| Science instruments | 16.2 | 1.3 | 21.1 |
| Battery + PDS | 0.4 | 1.3 | 0.5 |
| Communication | 3.7 | 1.3 | 4.8 |
| Thermal | 4.2 | 1.3 | 5.5 |
| C&DH | 3.1 | 1.3 | 4.0 |
| **Total** | **34.2** | | **44.4** |

Typical dimensions for the mass analyzer component of the LDMS are around 12–15 cm in height and 6 cm in diameter [54,56,57]. We assume that the fully developed instrument will fit within a 25 × 25 × 20 cm box—its exact dimensions depend largely on the design of the sample delivery system, which is not well determined at present. The total mass of the instrument may range from 11–15 kg, again depending on the design of the delivery system. We present a notional design for the LDMS descent probe in Figure 5.

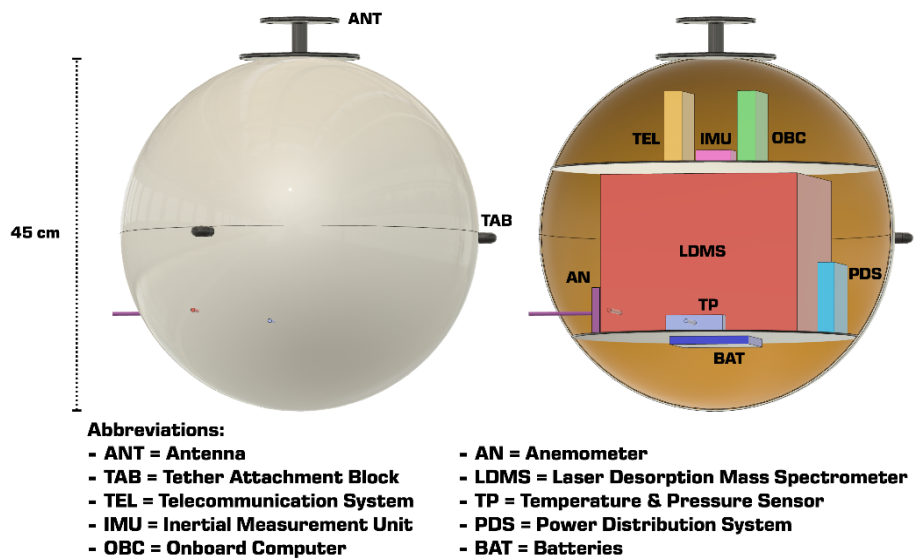

**Figure 5.** Concept design for the LDMS descent probe.

The probe is a 45 cm spherical titanium shell of 2 mm thickness. The forward and aft shelves are made of beryllium of thickness equal to that of the hull wall and are coated with sodium silicate. A thin layer of Kapton lines the interior of the probe. The LDMS is fastened to the forward shelf, with other non-science subsystems distributed among both shelves. There are three protrusions in the hull wall: one for the anemometer; a second for the temperature and pressure sensor; and a third for the antenna. A single penetration is required for the LDMS sample collection mechanism. The total mass of the probe likely will not exceed 44.4 kg, with the current best estimate being 34.2 kg.

## 5. Summary and Conclusions

The VLF Habitability Mission is a streamlined mission concept designed to answer key questions about the habitability of the clouds of Venus as well as search for signs of life. Determining cloud particle characteristics—their acidity and chemical composition, including water content and the presence of dissolved metals—is of central focus. The mission also aims to resolve longstanding chemical anomalies of the clouds of Venus, including the presence of ammonia, molecular oxygen, and the anomalous shapes of the Mode 3 cloud particles, all of which are motivated by the search for signs of habitability and life.

The data on cloud particles' physical and chemical properties returned by the Habitability Mission will also inform the future development of sample capture, storage, and transfer technologies. Together with balloon deployment and operation, the mission will demonstrate critical technologies needed for a successful atmosphere sample return mission.

**Supplementary Materials:** Not applicable.

**Author Contributions:** Conceptualization, S.S., J.J.P., C.E.C., S.J.S., R.A., W.P.B. and D.H.G.; methodology, S.S., J.J.P., C.E.C., S.J.S., R.A., W.P.B. and I.I.; formal analysis, S.S., J.J.P., C.E.C., S.J.S., R.A., W.P.B. and I.I.; investigation, S.S., J.J.P., C.E.C., S.J.S., R.A., W.P.B., D.H.G., M.U.W., P.K., S.P.W., I.I., M.P. and L.K.; writing—original draft preparation, S.S. and J.J.P.; writing—review and editing, S.S., J.J.P., C.E.C., S.J.S., R.A., W.P.B., D.H.G., M.U.W., P.K., S.P.W., I.I., M.P. and L.K. All authors have read and agreed to the published version of the manuscript.

**Funding:** This research was partially funded by Breakthrough Initiatives, the Change Happens Foundation, and the Massachusetts Institute of Technology.

**Institutional Review Board Statement:** Not applicable.

**Informed Consent Statement:** Not applicable.

**Data Availability Statement:** Not applicable.

**Acknowledgments:** We thank the extended Venus Life Finder Mission team for useful discussions. List of the individuals involved as the VLF extended Venus Life Finder Mission team can be found here: https://venuscloudlife.com/, accessed on 14th of May 2022

**Conflicts of Interest:** The authors declare no conflict of interest.